\documentclass[iop]{emulateapj}
\usepackage{apjfonts}
\usepackage{natbib}
\usepackage{amsfonts}
\usepackage{amsmath}
\usepackage[usenames]{color}
\usepackage{ulem}
\usepackage{verbatim}
\usepackage{longtable}
\usepackage{array}
\usepackage{epsfig}
\usepackage{graphicx}
\usepackage{threeparttable}
\newcommand\msun{M$_\odot$}
\newcommand\mzams{$M_{\rm ZAMS}$}

\shorttitle{The Supernova Progenitor Mass Distributions of M31 and M33}
\shortauthors{Jennings et al.}

\begin{document}

\title{The Supernova Progenitor Mass Distributions of M31 and M33: Further Evidence for an Upper Mass Limit}

\author{Zachary G. Jennings\altaffilmark{1}}
\author{Benjamin F. Williams\altaffilmark{2}}
\author{Jeremiah W. Murphy\altaffilmark{4}}
\author{Julianne J. Dalcanton\altaffilmark{2}}
\author{Karoline M. Gilbert\altaffilmark{2,3}}
\author{Andrew E. Dolphin\altaffilmark{5}}
\author{Daniel R. Weisz\altaffilmark{1,2,6}}
\author{Morgan Fouesneau\altaffilmark{2}}

\altaffiltext{1}{University of California Observatories, Santa Cruz, CA 95064, USA; zgjennin@ucsc.edu}
\altaffiltext{2}{Box 351580, The University of Washington
  Seattle, WA 98195}
\altaffiltext{3}{Space Telescope Science Institute, Baltimore, MD 21218, USA}
\altaffiltext{4}{Department of Physics, Florida State University, Tallahassee, FL 32306}
\altaffiltext{5}{Raytheon, 1151 E. Hermans Road, Tucson, AZ 85706; adolphin@raytheon.com}
\altaffiltext{6}{Hubble Fellow}

\begin{abstract}
Using Hubble Space Telescope (\textit{HST}) photometry to measure star
formation histories, we age-date the stellar populations surrounding
supernova remnants (SNRs) in M31 and M33. We then apply stellar
evolution models to the ages to infer the corresponding masses for
their supernova progenitor stars.  We analyze 33 M33 SNR progenitors
and 29 M31 SNR progenitors in this work. We then combine these
measurements with 53 previously published M31 SNR progenitor
measurements to bring our total number of progenitor mass estimates to
115. To quantify the mass distributions, we fit power laws of the form
$dN/dM \propto M^{-\alpha}$. Our new larger sample of M31 progenitors
follows a distribution with $\alpha=4.4^{+0.4}_{-0.4}$, and the M33
sample follows a distribution with $\alpha=3.8^{+0.4}_{-0.5}$. Thus
both samples are consistent within the uncertainties, and the full
sample across both galaxies gives $\alpha=4.2^{+0.3}_{-0.3}$.  Both
the individual and full distributions display a paucity of massive
stars when compared to a Salpeter initial mass function (IMF), which
we would expect to observe if all massive stars exploded as SN that
leave behind observable SNR. If we instead fix $\alpha=2.35$ and treat
the maximum mass as a free parameter, we find $M_{\rm
max}{\sim}35-45$ \msun, indicative of a potential maximum cutoff mass
for SN production.  Our results suggest that either SNR surveys are
biased against finding objects in the youngest ($<$10~Myr old)
regions, or the highest mass stars do not produce SNe.

\end{abstract}


\section{INTRODUCTION}

While core-collapse supernovae (CCSNe) are linked both theoretically
and observationally to the deaths of massive stars, the exact mapping
between the mass of the progenitor and the nature of its death is an
unresolved question. It is yet unknown exactly how the properties of a
given star, specifically its mass, will affect the eventual SN type.
Archival imaging of progenitor stars has been successful at linking
Type II-P SNe to red supergiant (RSG) stars. However, it is difficult
to predict the fate of more massive stars and the progenitors of
uncommon types SN types due to their rarity. Significant questions
exist as to which mass ranges result in which types of SN, to what
extent other properties such as metallicity affect this range, and
whether all massive stars actually explode as SN.

\citet{smartt2009} identified what they termed the "red supergiant
problem"--- an observed lack of progenitors between 18 and 30 \msun
which would be expected to explode as Type II SNe. They postulated
that massive stars in this range may fail to explode as SNe, ending
their lives in some other way. \citet{horiuchi2011} compared measured
massive star formation rates with the measured CCSN rate and found
that twice as many massive stars are formed than explode as CCSNe.
They explored a wide variety of explanations for this observation, and
found that measurement errors on either rate could not solely explain
the discrepancy.  \citet{smith2011} examined observed rates of
different SN types and found it impossible to explain the observed
rates of more exotic SNe using solely single-star
evolution. \citet{kochanek2014} explored the observed compact remnant
mass function and found that it could be well-explained by a failed-SN
scenario, where some massive stars did not actually produce SNe.
Together, these recent studies all present evidence suggesting that a
model in which all massive stars end their lives as SNe through
single-star evolutionary channels is an incomplete picture.

Having a greater number of progenitor mass measurements naturally
makes these issues easier to address. Unfortunately,
direct imaging has requirements that limit the frequency with which it
may be applied.  Since one must directly observe the SN to identify
the progenitor in pre-explosion imaging, one is clearly limited by the
SN rate in the local universe. In addition, archival imaging of
sufficient depth must exist, nearly always necessitating that the
field in question has been previously imaged with either the Advanced
Camera for Surveys (ACS), Wide-Field Planetary Camera 2 (WFPC-2), or
Wide-Field Camera 3 (WFC3) instruments aboard \textit{Hubble Space
Telescope} (\textit{HST}). As a result of these
limitations, only $\sim$25 SN progenitors currently have mass
constraints in the literature, and around half of these are only upper
limits \citep{smartt2002,vandyk2003a,
vandyk2003b,smartt2004,maund2005,hendry2006,li2005,li2006,li2007,
smartt2009,smartt2009b,galyam2007,galyam2009,smith2011b,maund2011,vandyk2011,
fraser2012,vandyk2012b,vandyk2012a,fraser2014,maund2014b,maund2014}.

\begin{figure*}
\centering
%\plottwo{m31_paper2.ps}{m33_paper2.ps}
\begin{minipage}[b]{.45\textwidth}
\centering
\includegraphics[width=\linewidth]{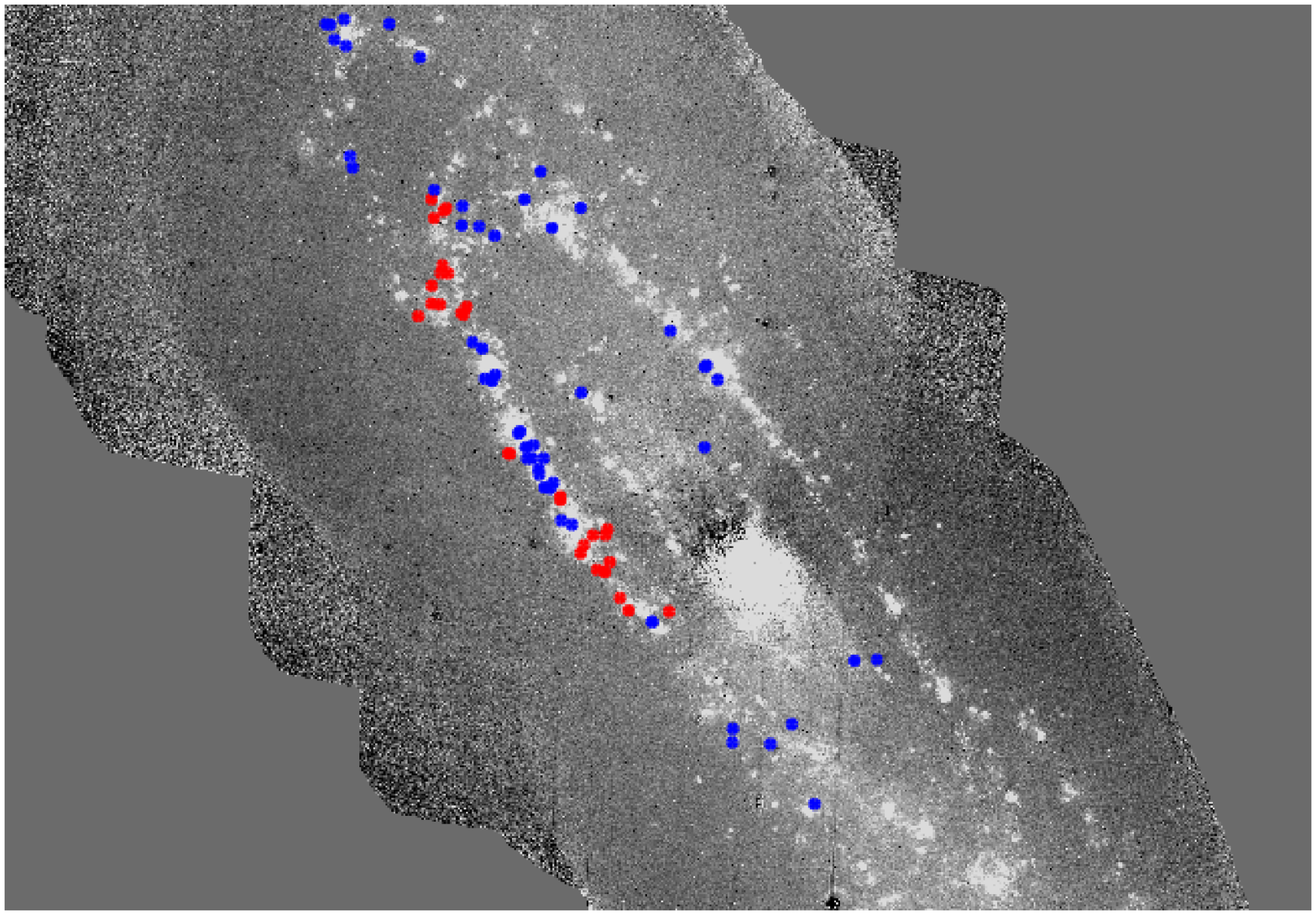}
\end{minipage}%
\hskip 0.3cm
\begin{minipage}[b]{.45\textwidth}
\centering
\includegraphics[width=\linewidth]{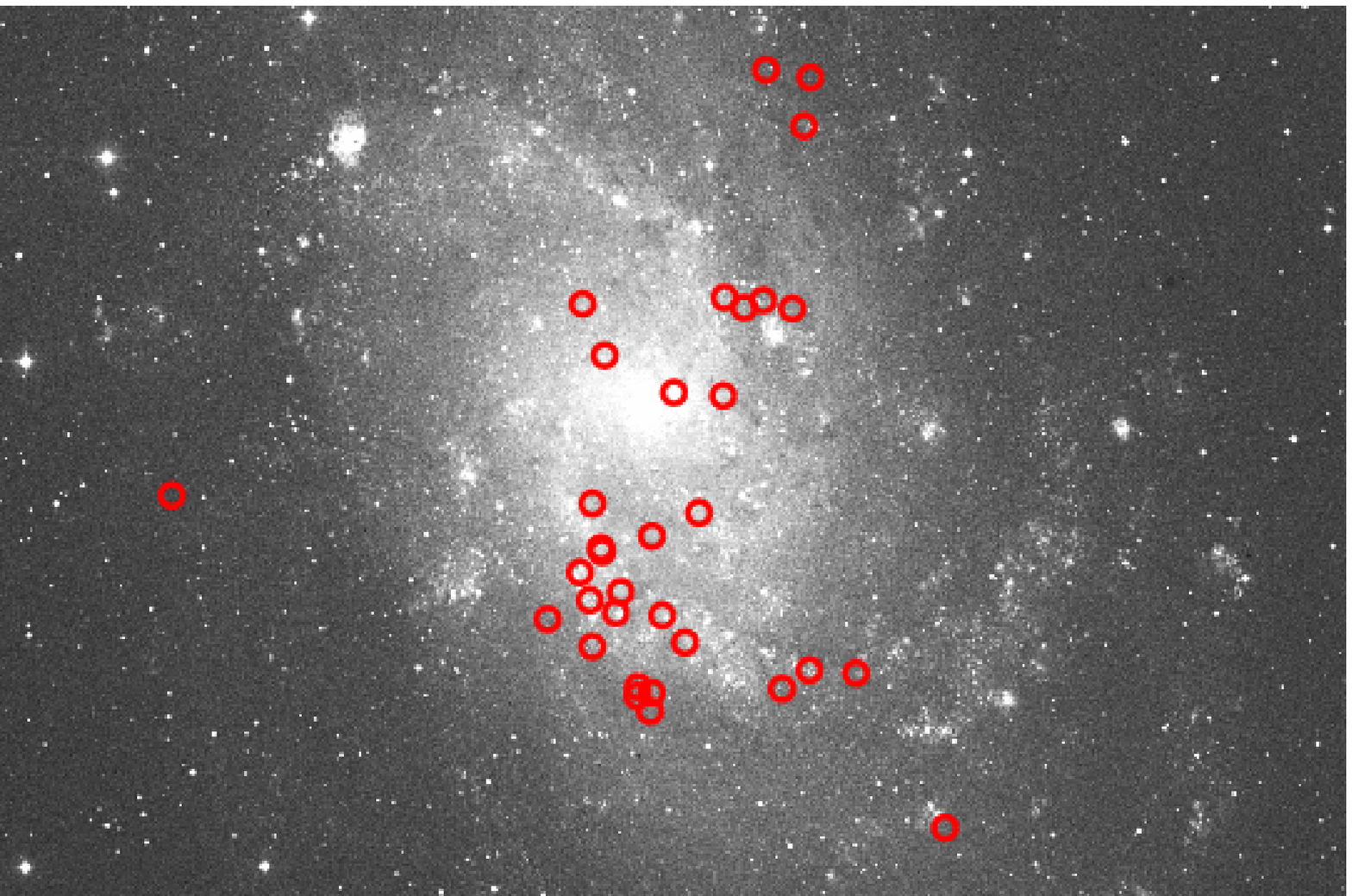}
\end{minipage}
\caption{Left Panel: Locations of the combined M31 SNR sample
superposed on a star-subtracted H$\alpha$ image of M31 from
\citet{williams1995}.  SNRs from J12 are plotted in blue, while those
from this paper are plotted in red. As the new M31 SNR all come from
recently acquired PHAT survey data, they are highly correlated
spatially. Right Panel: Locations of M33 SNR analyzed in this paper on
a DSS image of M33.}
\label{spatial_snr}
\end{figure*}

In this work, we employ stellar population analysis as an alternative
means of constraining the progenitor mass. The literature features
many examples of using the age of a surrounding stellar population to
constrain the progenitor stars of directly observed transients
\citep{efremov1991,
walborn1993,panagia2000,barth1996,vandyk1999,maiz2004,wang2005,
vinko2009,crockett2008, gogarten2009, murphy2011}. Stellar population
analysis offers a means of constraining progenitor masses in cases
where pre-explosion direct imaging does not exist, as well as a
complementary constraint in cases where pre-explosion images of the
progenitor are available.

Since one does not need to identify the specific progenitor star to
determine the likely age of the star formation (SF) event that
produced the progenitor, one may apply stellar population analysis to
the locations of cataloged supernova remnants (SNRs), drastically
increasing the number of potential progenitor measurements. While SNe
are typically only visible on time scales of ${\sim}10^2$ days, SNR
remain observable for ${\sim}10^4$ years. Assuming a SNe rate of
${\sim}10^{-2}$ yr$^{-1}$ galaxy$^{-1}$ \citep{cappellaro1999}, one
would naively expect to be able to make $\sim$100 progenitor mass
estimates per galaxy, provided they are close enough for resolved
stellar population analysis.  While this approach naturally prevents
direct investigation of the SN type, it offers in exchange the
leverage to investigate more potential targets.

Only a few studies have explored this application so
far. \citet{badenes2009} used the star formation maps of
\citet{harris2009} to estimate progenitor ages and masses for 8 SNR in
the LMC.  In \citet{jennings2012} (hereafter J12), we identified
\textit{HST}-observed fields coincident with cataloged SNR in M31 and
analyzed resolved stellar photometry of the surrounding stellar
population.  We created color-magnitude diagrams (CMDs) of these
populations and fit them to recover the star formation histories
(SFHs). By identifying coeval populations in the SFH, we assigned
likely ages and therefore likely masses to the SN progenitors.  Our
final result consisted of mass estimates for 53 likely CCSN
progenitors, which we presented as a mass distribution.  The
primary finding in J12 was a paucity of massive stars in the recovered
mass distribution when compared to a Salpeter initial mass function
(IMF) \citep{salpeter1955}. This suggests that some fraction of
massive stars are not exploding as SNe which leave behind observable
SNR, in agreement with the evidence found in the above studies.

This paper represents several extensions to the work presented in J12.
First, we more than double the number of total progenitor star mass
estimates, improving the statistics on our mass distribution
fitting. We analyze 29 additional SNR in M31 for which we have
recently acquired new Panchromatic Hubble Andromeda Treasury
(PHAT,~\citealt{dalcanton2012}) data. Second, we have expanded
our sample to M33, an environment with roughly half the
metallicity of M31 \citep{barker2011}, by analyzing 33 additional M33 SNR
from various archival \textit{HST} datasets. We also verify our result
against potential biases due to host galaxy inclination or type.
Finally, we update our mass distribution fitting method to a
probabilistic framework, based on Markov Chain Monte Carlo (MCMC)
techniques, in which we reliably include our mass uncertainties.

An outline of the paper is as follows: In \S2 we briefly summarize our
technique of mass estimation and discuss the SNR catalogs used in this
study.  In \S3, we present our mass estimations for the additional
regions measured in this study.  In \S4 we discuss the mass
distributions of our SNR progenitors.  Finally, in \S5, we summarize
the main conclusions of the paper.

\section{METHODS}
In this section, we provide a brief summary of the main points of our
methodology; the full details of our technique are presented in J12.

\subsection{Data Selection and Photometry}
First, we select SNRs from catalogs which overlap with \textit{HST}
fields in M31 and M33.  We limited our search to fields observed with
the ACS and WFPC-2 instruments, as the optical CMDs tend to be deeper
and offer the best SFH constraints. For M31, we combine three SNR
catalogs to select targets from:
\citet{braun1993,magnier1995,williams1995}. All three catalogs make
use of [S~II]-to-H$\alpha$ ratios to identify SNR. Our M33 SNR catalog
comes from \citet{long2010}, who selected SNRs using both
[S~II]-to-H$\alpha$ ratios and X-ray observations.  \citet{long2010}
incorporated all previous SNR catalogs from the literature in their
analysis.  As mentioned in \S1, all new M31 data analyzed in this work
come from newly available PHAT datasets and therefore have uniform
filters and exposure times, while the M33 data comes from various
archival datasets coincident with SNR.  The locations of SNR analyzed
in this paper and J12 are plotted in Fig.~1. The full list of SNR
analyzed, including the associated \textit{HST} datasets used, are in
Table 1.

We used the photometry pipeline from the ACS Nearby Galaxy Treasury
Program \citep{dalcanton2009} and the PHAT Program
\citep{dalcanton2012} to perform resolved stellar photometry on all
stars in the selected fields.  The full details of how this photometry
is performed are provided in
\citet{dalcanton2009,dalcanton2012}. Briefly, the pipeline uses
DOLPHOT \citep{dolphin2000} to fit the well-characterized ACS
point-spread function to all of the point sources in the
image. Photometric measurements are then converted to Vega magnitudes
using zero-points from the ACS handbook. We use fake star tests to
assess photometric uncertainties and completeness; $10^5$ fake stars
of known color and magnitude are inserted into the full fields and
blindly recovered using the identical software.

We assume distance moduli of 24.47 to M31 \citep{mcconnachie2005} and
of 24.69 to M33 \citep{barker2011}. Typical uncertainties of from these surveys are
$\sim$0.05 mag for M31 and $\sim$0.1 mag for M33. We fix the distance moduli to
these values for the remainder of the paper.

We found in J12 that it was
difficult to constrain older SF events in shallow fields. We defined the "depth" of the field
as the point at which photometric completeness dropped below 50\% and adopted
the same depth cut as J12 for this work. The values used are
$F475W=24.5$ for M31 data, shifted to
$F475W=24.7$ for the M33 fields.
In practice, nearly all the fields examined in this work have depths significantly below
these depth limits, making the precise placement of the cut largely unimportant.

\begin{table*}
\centering
\begin{threeparttable}
\begin{tabular}{c c c c c c c c}
\caption{Sample SNR for M31 and M33} \\
\hline
SNR ID $^{\rm \bf A}$ & R.A. (J2000) & Dec. (J2000) & \textit{HST} Field $^{\rm \bf B}$ &
\textit{HST} Program ID & 
Instrument & Filters \& 50\% Completeness Limits & Catalog $^{\rm \bf C}$ \\
 & (deg.) & (deg.) &  & &  & \\
\multicolumn{8}{c}{M31 SNR} \\
\hline
2-038 & 10.9933 & 41.223 & B02F04 & 12073 & ACS/WFC & F475W=27.06, F814W=25.74 & 1 \\
2-039 & 11.0173 & 41.2853 & B04F10 & 12107 & ACS/WFC & F475W=27.23, F814W=25.82 & 1 \\
2-040 & 11.0228 & 41.3423 & B04F05 & 12107 & ACS/WFC & F475W=27.21, F814W=25.69 & 1 \\
2-041 & 11.0285 & 41.2678 & B04F10 & 12107 & ACS/WFC & F475W=27.11, F814W=25.84 & 1 \\
2-042 & 11.0463 & 41.2721 & B04F10 & 12107 & ACS/WFC & F475W=27.47, F814W=26.11 & 1 \\
2-043 & 11.0848 & 41.3002 & B04F04 & 12107 & ACS/WFC & F475W=27.43, F814W=26.09 & 1 \\
2-052 & 11.4151 & 41.7867 & B14F04 & 12072 & ACS/WFC & F475W=27.39, F814W=26.21 & 1 \\
2-053 & 11.4313 & 41.8822 & B16F05 & 12106 & ACS/WFC & F475W=27.67, F814W=26.37 & 1 \\
2-054 & 11.4661 & 41.7114 & B14F15 & 12072 & ACS/WFC & F475W=27.56, F814W=26.24 & 1 \\
BW45 & 10.8796 & 41.1996 & B02F11 & 12073 & ACS/WFC & F475W=26.78, F814W=25.33 & 2 \\
BW51 & 10.9729 & 41.2011 & B02F04 & 12073 & ACS/WFC & F475W=27.23, F814W=25.96 & 2 \\
BW58 & 11.0562 & 41.3319 & B04F04 & 12107 & ACS/WFC & F475W=27.42, F814W=25.92 & 2 \\
BW59 & 11.0775 & 41.3146 & B04F04 & 12107 & ACS/WFC & F475W=27.46, F814W=25.98 & 2 \\
BW63 & 11.1321 & 41.3933 & B06F10 & 12105 & ACS/WFC & F475W=27.44, F814W=26.1 & 2 \\
BW64 & 11.1325 & 41.3986 & B06F10 & 12105 & ACS/WFC & F475W=27.44, F814W=26.1 & 2 \\
BW79 & 11.2542 & 41.4734 & B08F03 & 12075 & ACS/WFC & F475W=27.29, F814W=26.11 & 2 \\
BW94 & 11.4029 & 41.8997 & B16F05 & 12106 & ACS/WFC & F475W=27.52, F814W=26.42 & 2 \\
BW95 & 11.4083 & 41.8955 & B16F05 & 12106 & ACS/WFC & F475W=27.54, F814W=26.42 & 2 \\
BW96 & 11.4104 & 41.7999 & B14F04 & 12072 & ACS/WFC & F475W=27.29, F814W=26.18 & 2 \\
BW98 & 11.4154 & 41.7316 & B14F10 & 12072 & ACS/WFC & F475W=27.43, F814W=26.04 & 2 \\
BW100 & 11.435 & 41.7648 & B14F10 & 12072 & ACS/WFC & F475W=27.75, F814W=26.46 & 2 \\
BW101 & 11.4354 & 41.7331 & B14F09 & 12072 & ACS/WFC & F475W=27.62, F814W=26.39 & 2 \\
K293 & 11.0277 & 41.3325 & B04F05 & 12107 & ACS/WFC & F475W=27.1, F814W=25.68 & 3 \\
K413 & 11.1076 & 41.4136 & B06F04 & 12105 & ACS/WFC & F475W=27.25, F814W=25.82 & 3 \\
K638 & 11.2495 & 41.4728 & B08F03 & 12075 & ACS/WFC & F475W=27.37, F814W=26.11 & 3 \\
K763 & 11.3537 & 41.7284 & B14F17 & 12072 & ACS/WFC & F475W=27.35, F814W=25.96 & 3 \\
K774 & 11.3607 & 41.7147 & B12F05 & 12071 & ACS/WFC & F475W=27.54, F814W=26.2 & 3 \\
K782 & 11.3653 & 41.7175 & B12F05 & 12071 & ACS/WFC & F475W=27.54, F814W=26.2 & 3 \\
K817 & 11.3973 & 41.7863 & B14F04 & 12072 & ACS/WFC & F475W=27.42, F814W=26.11 & 3 \\
K854A & 11.4375 & 41.9152 & B16F05 & 12106 & ACS/WFC & F475W=27.65, F814W=26.34 & 3 \\
\hline 
SNR ID $^{\rm \bf A}$ & R.A. (J2000) & Dec. (J2000) & \textit{HST} Field $^{\rm \bf B}$
& \textit{HST} Program ID & 
Instrument & Filters \& 50\% Completeness Limits & Catalog $^{\rm \bf C}$ \\
 & (deg.) & (deg.) &  & &  & \\
\multicolumn{8}{c}{M33 SNR} \\ 
\hline
20 & 23.2874 & 30.4497 & ANY & 10190 & ACS/WFC & F606W=27.14, F814W=26.49 & 4 \\
30 & 23.3402 & 30.5253 & ANY & 10190 & ACS/WFC & F606W=27.03, F814W=26.31 & 4 \\
34 & 23.367 & 30.5264 & ANY & 10190 & ACS/WFC & F606W=26.71, F814W=26.12 & 4 \\
37 & 23.3727 & 30.8197 & H10 & 5914 & WFPC2 & F555W=26.73, F814W=25.67 & 4 \\
38 & 23.3759 & 30.7955 & H10 & 5914 & WFPC2 & F555W=26.5, F814W=25.36 & 4 \\
40 & 23.3806 & 30.7051 & ANY & 9873 & WFPC2 & F606W=25.36, F814W=25.31 & 4 \\
41 & 23.3825 & 30.517 & ANY & 10190 & ACS/WFC & F606W=26.76, F814W=26.18 & 4 \\
43 & 23.3975 & 30.7089 & ANY & 9873 & WFPC2 & F606W=25.69, F814W=24.67 & 4 \\
44 & 23.3984 & 30.8231 & H10 & 5914 & WFPC2 & F555W=27.03, F814W=25.84 & 4 \\
48 & 23.4084 & 30.7051 & ANY & 9873 & WFPC2 & F606W=26.14, F814W=24.89 & 4 \\
49 & 23.4194 & 30.6613 & ANY & 10190 & ACS/WFC & F606W=25.52, F814W=24.77 & 4 \\
50 & 23.4197 & 30.7099 & ANY & 9873 & WFPC2 & F606W=25.13, F814W=24.09 & 4 \\
57 & 23.4321 & 30.6032 & DISK1 & 10190 & ACS/WFC & F475W=26.93, F814W=25.75 & 4 \\
58 & 23.4386 & 30.5389 & SRV6 & 6640 & WFPC2 & F555W=25.59, F814W=24.44 & 4 \\
59 & 23.4477 & 30.6624 & ANY & 10190 & ACS/WFC & F606W=24.89, F814W=24.52 & 4 \\
61 & 23.4521 & 30.5522 & SRV6 & 6640 & WFPC2 & F555W=25.48, F814W=24.29 & 4 \\
62 & 23.4573 & 30.5138 & SRV6 & 6640 & WFPC2 & F555W=25.91, F814W=24.75 & 4 \\
63 & 23.4579 & 30.5046 & H38 & 5914 & WFPC2 & F555W=26.1, F814W=24.89 & 4 \\
64 & 23.4588 & 30.5913 & DISK1 & 10190 & ACS/WFC & F475W=26.86, F814W=25.71 & 4 \\
66 & 23.4653 & 30.5166 & SRV6 & 6640 & WFPC2 & F555W=25.93, F814W=24.77 & 4 \\
67 & 23.4655 & 30.5121 & SRV7 & 6640 & WFPC2 & F555W=26.27, F814W=25.06 & 4 \\
69 & 23.4762 & 30.5633 & ANY & 10190 & ACS/WFC & F606W=25.61, F814W=24.52 & 4 \\
71 & 23.4788 & 30.5531 & ANY & 10190 & ACS/WFC & F606W=25.78, F814W=25.01 & 4 \\
74 & 23.4874 & 30.583 & DISK1 & 10190 & ACS/WFC & F475W=26.85, F814W=25.75 & 4 \\
75 & 23.488 & 30.6801 & R14 & 5914 & WFPC2 & F555W=24.77, F814W=23.53 & 4 \\
76 & 23.488 & 30.585 & DISK1 & 10190 & ACS/WFC & F475W=26.82, F814W=25.75 & 4 \\
77 & 23.4919 & 30.536 & ANY & 10190 & ACS/WFC & F606W=26.06, F814W=25.4 & 4 \\
80 & 23.4934 & 30.6068 & DISK1 & 10190 & ACS/WFC & F475W=26.81, F814W=25.69 & 4 \\
81 & 23.4938 & 30.559 & ANY & 10190 & ACS/WFC & F606W=25.38, F814W=24.99 & 4 \\
83 & 23.4997 & 30.5726 & ANY & 10190 & ACS/WFC & F606W=25.61, F814W=24.78 & 4 \\
84 & 23.5013 & 30.7054 & R14 & 5914 & WFPC2 & F555W=24.96, F814W=23.78 & 4 \\
91 & 23.5177 & 30.5492 & ANY & 10190 & ACS/WFC & F606W=25.88, F814W=25.03 & 4 \\
134 & 23.7352 & 30.6064 & ANY & 9873 & WFPC2 & F606W=27.61, F814W=26.46 & 4 \\
\hline
\end{tabular}
\begin{tablenotes}[para,flushleft]
\textbf{A:} IDs taken from respective catalogs.
\textbf{B:} Field designation as listed in MAST.
\textbf{C:} Catalogs 1, 2, 3, and 4 are \citet{magnier1995}, \citet{williams1995}, \citet{braun1993}, and
\citet{long2010} respectively. 
\end{tablenotes}
\end{threeparttable}
\end{table*}

\subsection{SFH Measurement and Age Determination} 

After performing full-field photometry, we select all stars in an
annulus of $\sim50$ pc around the location of the SNR of interest,
which we use to then measure the SFH of the region.  We use the
software package MATCH \citep{dolphin2002} to fit the observed CMD
with synthetic ones based on the models of
\citet{marigo2008,girardi2010}. For purposes of populating the models,
we assume a Salpeter IMF ($dN/dM\propto M^{-2.35}$) and a binary
fraction of 0.35.  J12 and \citet{gogarten2009} found no appreciable
effect on measured SFH values from reasonable variations of these
parameters. MATCH also makes use of fake stars to evaluate
completeness and photometric uncertainties, which we extract from an
annulus $\sim2.5$ times the size to ensure a sufficient number of fake
stars. The distance modulus is fixed to the respective value for each
galaxy from above. The CMD is binned with bin sizes of 0.3 in magnitude and 0.15 in
color.  Metallicity is constrained to have a spread of $\sim$0.15 dex,
and to increase or stay constant with time.

Our age bins for SFH determination are in log-space from $\log({\rm
Age})=6.60$ (4 Myr) to $\log({\rm Age})=10.10$ (12.5 Gyr) in steps of
0.05 dex. For a solar metallicity system, these ages would correspond to supernova progenitors
of roughly 52~\msun~and 7.3~\msun~respectively.
Note that any SF MATCH finds for ages younger than 4 Myr is
included in the 4 Myr to 4.4 Myr bin. Thus SF found in this youngest bin
actually acts as an upper limit on the age, constraining the SF to be
younger than 4.4 Myr. The final result from our CMD fitting analysis
is a SFH from the present back to 12.5 Gyr.

Our treatment of reddening in our CMD fitting was the same as in J12,
and merits further discussion. Our prescription for reddening is to
assume that $A_{\rm V}$ follows a top-hat distribution, the width of
which is specified by a $dA_{\rm V}$ parameter defined by the
user. MATCH then fits for the minimum value of this $A_{\rm V}$
distribution, defining the top-hat in the course of the fitting
procedure. To determine the value of this parameter most appropriate
for each SNR location, we increased the width of the $dA_{\rm V}$
parameter until the fit returned a reddening value consistent with the
\citet{schlegel98} foreground reddening value. In other words, we treat all
reddening in addition to foreground MW extinction as broadening of the
CMD due to differential reddening.

\subsection{Age/Mass Conversion}
For purposes of age/mass conversion, we use the same models as those
fitted to the CMDs.  For each age bin, we assume the star with the largest
zero-age main sequence mass (\mzams) remaining in the isochrone will be the next one to become a
SN. This mass is then taken as the progenitor mass for that given age bin.
We perform a simple linear interpolation across
age bins to get an estimated progenitor mass for an arbitrary age.  We
adopt the $Z=0.019$ isochrones for M31 and the $Z=0.008$ isochrones for
M33, which are consistent with gas-phase metallicity measurements for
the two galaxies \citep{blair1982,barker2011}. However, as
demonstrated in J12, the choice of metallicity is largely unimportant
in converting age to mass as the \mzams~masses are very
similar across reasonable ranges of metallicity. For a given age, the
difference in mass between the two isochrones is typically
a few percent.

Numerous sources of theoretical and observational evidence point
towards a minimum mass necessary for a star to undergo core
collapse. We wish to identify the age range where a population may
still produce a SN.  Coeval populations that are too old will have no
stars remaining massive enough to become core-collapse SNe.  In J12,
we found our data most consistent with a minimum mass of 7.3~\msun,
corresponding to an age of 50 Myr. We verified that our new data are
also consistent with this minimum mass, and therefore only consider ages
$\leq50$~Myr in our age distributions. However, we were not able to
improve this constraint using our current dataset.

We do not assume
any maximum mass at this stage. However, for ages younger than 4.4
Myr, we may only quote a limit on the progenitor mass of $>52$ \msun~
for the Z=0.019 isochrone (i.e. M31) and $>55$ \msun~for the Z=0.008 isochrone
(i.e. M33) because the optical CMDs are degenerate for ages younger than 4.4 Myr.

To estimate the age of the progenitor, we identify the age at which,
over the past 50 Myr, 50\% of the stars have formed. This median age
is then converted to a median mass using the interpolation defined
above. We also assign uncertainties to the age from two sources using
the methods discussed in the next section. Furthermore, we calculate
full probability distributions for each progenitor.  We assume that
the progenitor originates from any SF that has occurred in the region
over the past 50~Myr, where the fraction of stellar mass formed in a
given age bin corresponds to the probability that the progenitor star
was of that age.  Thus, all of our probability distributions sum to one
because they include all SF that has occurred over the past 50 Myr.

\subsection{Treatment of Uncertainties}
Uncertainties on the age estimate may arise either from uncertainties
in the CMD fitting process, or from the age spread present in the
stellar population.

To characterize uncertainties from fitting, we use a Monte Carlo (MC)
technique in which we resample the CMD to account for Poisson noise
using the MATCH software (see \citealt{dolphin2002}). We also apply random shifts to
the isochrones in temperature,
with $\sigma=0.02$, and bolometric luminosity, with $\sigma=0.17$,
during this procedure. These random shifts are intended to mimic
potential uncertainties in the stellar evolution models themselves.
The magnitude of the luminosity shift is also larger than the uncertainty
in the distance, thereby incorporating those uncertainties.
Each resampling is then
refit using our identical CMD fitting procedure, but with fixed
reddening. These resulting SFHs provide uncertainties on the SF in a
given bin. With these uncertainties, we calculate a probability
distribution for the value of the median age.  We adopt the 16\% and
84\% range of this distribution as our uncertainty in the
median age.

Another source of uncertainty is the intrinsic spread of the SFH
across multiple age bins. There are many SNR populations in which
there are two (or more) distinct SF events, and certainly some where
a single SF event may spread across two or more adjacent bins. To account for this, we
find the locations where 16\% and 84\% of the stellar mass was formed
and adopt these as our uncertainties due to the spread of SF.  These
age-spread uncertainties therefore include an age range that contains
68\% of the young stellar mass present in the region.

We add these two sources of uncertainty in quadrature to find our
final (potentially asymmetric) estimates of uncertainty on the median
age. Finally, we argue that we cannot truly constrain our progenitors
to any better precision that that afforded by the age bins in which
we've performed our CMD fitting. In other words, we always round our
uncertainties to the nearest age bin boundary.

In practice, the second source of uncertainty due to the spread of SF
across multiple age bins tends to dominate that provided by the random
uncertainties in the fitting process. This is because, while the
fitting process may affect the amount of SF found in a given bin, the
change in SF must be very large to modify the estimate of the \textit{median} age.
The random uncertainty must be enough to
significantly change the relative prominence of a preferred burst of
SF. For most of the MC realizations, it isn't, and as a
result, the spread of SF tends to be the largest factor.

The final source of uncertainty, which we neglect in determining our final answers,
is the conversion
from age to mass, which depends on the models applied. We do not
include this uncertainty in our mass tables, as our uncertainties
assume the \citet{marigo2008} and \citet{girardi2010} models to be
consistent with our SFH fitting as they are used in both processes.  In J12, we estimated that the
effects of applying different models for the age/mass conversion would
potentially be $\sim$0.5 - 1.0 \msun (younger ages have larger
uncertainty) by comparing the predicted masses of
\citet{pietrinferni2004} with those of \citet{marigo2008} and
\citet{girardi2010}. However, as we have used the \citet{marigo2008,girardi2010}
models to derive these ages in the first place, it is not clear that comparing
the \citet{pietrinferni2004} masses to those of our chosen isochrones is
reflective of the actual data.

\subsection{Assumptions in our Methodology}
We make multiple assumptions in our technique based upon
justifications detailed in J12.  In this section, we briefly review
these assumptions and their potential impacts on our conclusions.

First, naturally, in using the surrounding stellar population
to infer information about the progenitor star, we are assuming that
the two are evolutionarily linked. Given that the vast majority of
stars form in clusters \citep{lada2003} with coeval populations, and that
that these populations remain spatially associated on 50 pc scales for 50 Myr, we
expect this to be a reasonable assumption (see also
\citealt{bastian2006,gogarten2009b,eldridge2011}).

Related to this point is that our methodology is also contingent on
the accuracy, completeness, and selection effects of the SNR catalogs
used. If the SNR catalogs are biased towards one specific type of
progenitor or environment, this could bias the overall inferred distribution
\citet{braun1993} and \citet{magnier1995} both consider
extinction to be a minor problem. Both studies note that detection
biases may exist as a function of the age of the \textit{remnant}, but
don't note any biases towards any particular age of the surrounding
stellar population. Related to this, we have also assumed that SNRs
are similarly behaved in different environments. If SNRs in differently aged
populations had significantly different lifetimes or luminosities, we would naturally be subject
to selections effects from this.

We assume for our study that the SNR catalogs used
are not biased towards any particular progenitor age, and we have
included as many SNR locations as possible in our analysis.
We note that the consistency of the M33 and M31 results and the consistency
between different subsets of the M31 sample, which depend on different
SNR catalogs. This suggests that there are not significant selection biases
imposed by detection method or survey location. However,
we have no real way to test the second caveat, that SNRs from differently aged populations
have different lifetimes or detectability thresholds. We discuss this more in \S4.3.

Next, our method provides no information about the type of SN which
created the SNR. For CCSNe, this ambiguity doesn't affect the mass
estimation process, since all CCSNe are linked to the deaths of
massive stars and therefore young stellar populations (although the mass distribution
of progenitors as a function of SN type is certainly an interesting question, we are unable
to investigate this with our methodology). We note that there could be
contaminants from thermonuclear type Ia SNe.  However, type Ia SNe
will generally be associated with older stellar populations. The absence of any recent
SF from the CMD analysis will allow one to remove some fraction of these contaminants.  We
identified six such SNRs with zero recent SF in J12. In this work, we only find one such SNR in our new
M31 sample (K413) and zero such SNRs in our M33 sample.
The new M31 data analyzed in this work essentially all comes from the
star forming ring of M31 due to the distribution of the PHAT survey,
so it is not entirely surprising that nearly all fields analyzed return
significant recent SF and that nearly all of the SNe are CCSNe. The
result of all analyzed M33 SNR populations having young SF is also
likely due to the fact that M33 has a higher SF intensity than M31.

Typical fractions of type Ia SNe compared to all SNe are $\sim$25\%,
although such a figure is dependent on the galaxy type in question
\citep{li2011}. We would expect a higher fraction for M31 and a lower
fraction for M33 based on morphological distinction.
In addition, there are many reasons to assume that
this Ia fraction may actually be an upper limit, with a bias resulting
from the faintness of type CCSNe compared to type Ia SNe (see
discussions in \citealt{thompson2009,
horiuchi2009,horiuchi2011}). Finally, both \citet{braun1993} and
\citet{magnier1995} explicitly note that they will be significantly
more biased towards identifying CCSNe remnants over type Ia remnants
due to the spatial distribution of their surveys.  We do not include
the M31 bulge in our sample, where most of the Ia events would be expected to occur.
Thus we actually expect Ia SNe in our sample to be a significantly
smaller fraction than would be expected based on galaxy-wide SN rate
studies.

Furthermore, in J12, we demonstrated that our results are robust to inclusion of
additional mis-identified type Ia SNe by removing additional potential
contaminants even up to 25\%. We identified those SNR-locations with minimal GALEX FUV
flux and removed those objects from our sample. The resulting progenitor mass distribution
was essentially unchanged
(the 95\% confidence interval on the power-law exponent changed from $2.7<\alpha<4.4$ to
$2.6<\alpha<4.3$).
We conclude that, while there is
uncertainty associated with an individual progenitor mass measurement
due to the possibility of it being a type Ia SNe, the overall
distributions of progenitor masses are largely robust to the inclusion
of a few progenitors with random ages since the vast majority of our SNR
will be the result of CCSNe. In other words, while there are almost certainly some
type Ia contaminants masquerading as CCSNe in our sample, they will be a small number
and will not significantly modify the overall conclusions. 

Finally, our method is contingent on the accuracy of the stellar
evolution models used to populate our model CMDs.
Our CMD-derived SFHs include estimates of these model uncertainties
(see \S2.4), although we neglect them during our mass conversion procedure.
It is worth noting that studies of directly identified
progenitor stars suffer from similar uncertainties in that they must
fit the spectral energy distributions to derive a luminosity and
temperature and use stellar evolution models to infer the mass of
late-stage massive stars (e.g. \citealt{smartt2009}).  These are
thought to be the most uncertain stages of the model predictions. Our
approach makes use of the entire CMD, making our results less
sensitive to the details of the late-stage stellar evolution models.

\section{RESULTS}
Examples of the process for mass determinations of M31 SNRs can be
seen in J12.  Two examples of the fitting process applied to M33 SNRs,
20 and 34, are displayed in Fig.~2 and Fig.~3.  The left panel of each
plots the observed CMD in red points, with the best-fit model produced
by MATCH in grayscale in the background. The right panel of both
displays the cumulative fractional SFH over the past 50 Myr for each
SNR, as fit by MATCH. The green highlighted region represents the
68\% confidence interval for the age and mass estimates. The mass
value is taken from where the cumulative line crosses the 50\%
cumulative fraction (marked with a blue line), with the upper and lower values
given by the 68\% confidence interval.

We provide these estimates for the M31 SNR in Table~2 and the M33 SNR
in Table~3.  We also include the number of main sequence stars and the
total number of stars in the 50 pc region around the remnant, the
total stellar mass formed over the past 50 Myr, and the $dA_{\rm V}$
parameter applied when fitting the CMDs.

\begin{figure*}
\centering
\epsfig{file=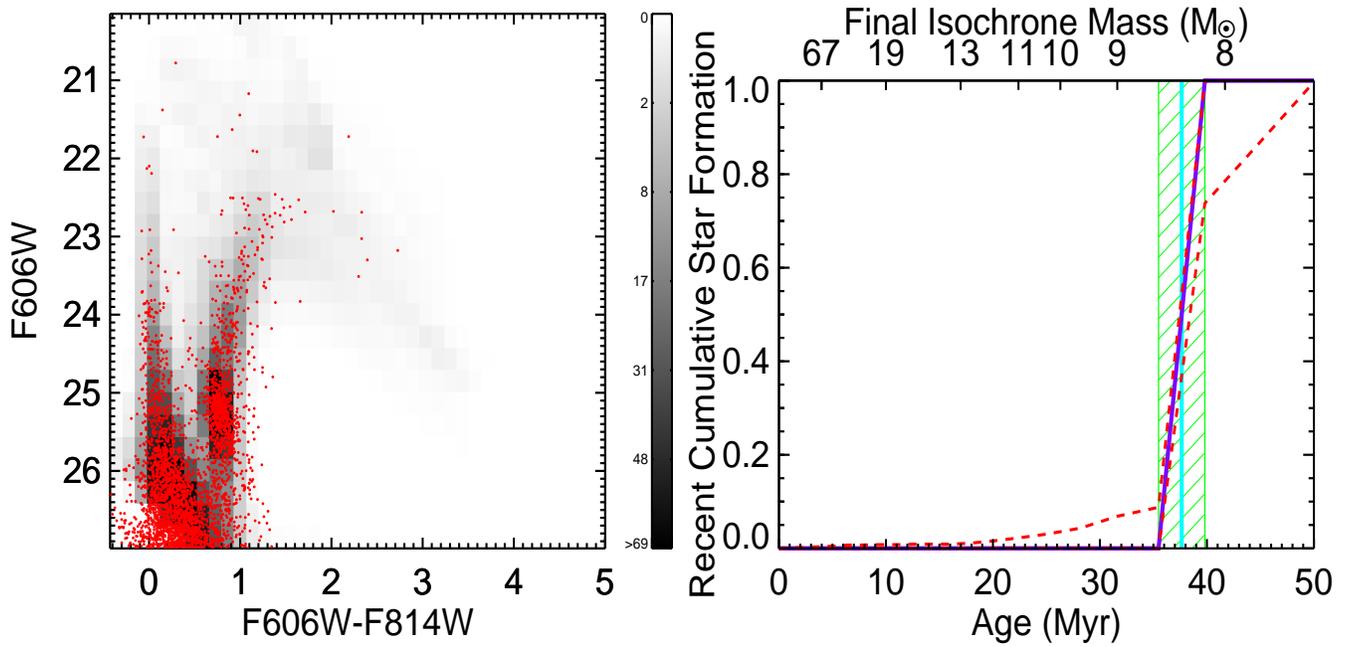,height=3.5in,width=7in}
\caption{Left Panel: Red points are observed F606W/F814W CMD for SNR
20 in M33. The background grayscale represents the best-fit model from
MATCH, with the stellar density scale on the right.  Right Panel:
Cumulative fractional SFH from the best-fit MATCH model over the most
recent 50 Myr.  The best fit SFH is the purple line. The orange
lines are 68\% uncertainties. The cross-hatched green region
represents the favored age and mass for this SNR progenitor, with
uncertainties calculated as described in \S2.4. The population
features one prominent burst of SF, favoring a well-constrained mass
of $8.6^{+0.3}_{-0.3}$~\msun. The blue line represents the median age.}
\label{mass_dist1}
\end{figure*}

\begin{figure*}
\centering
\epsfig{file=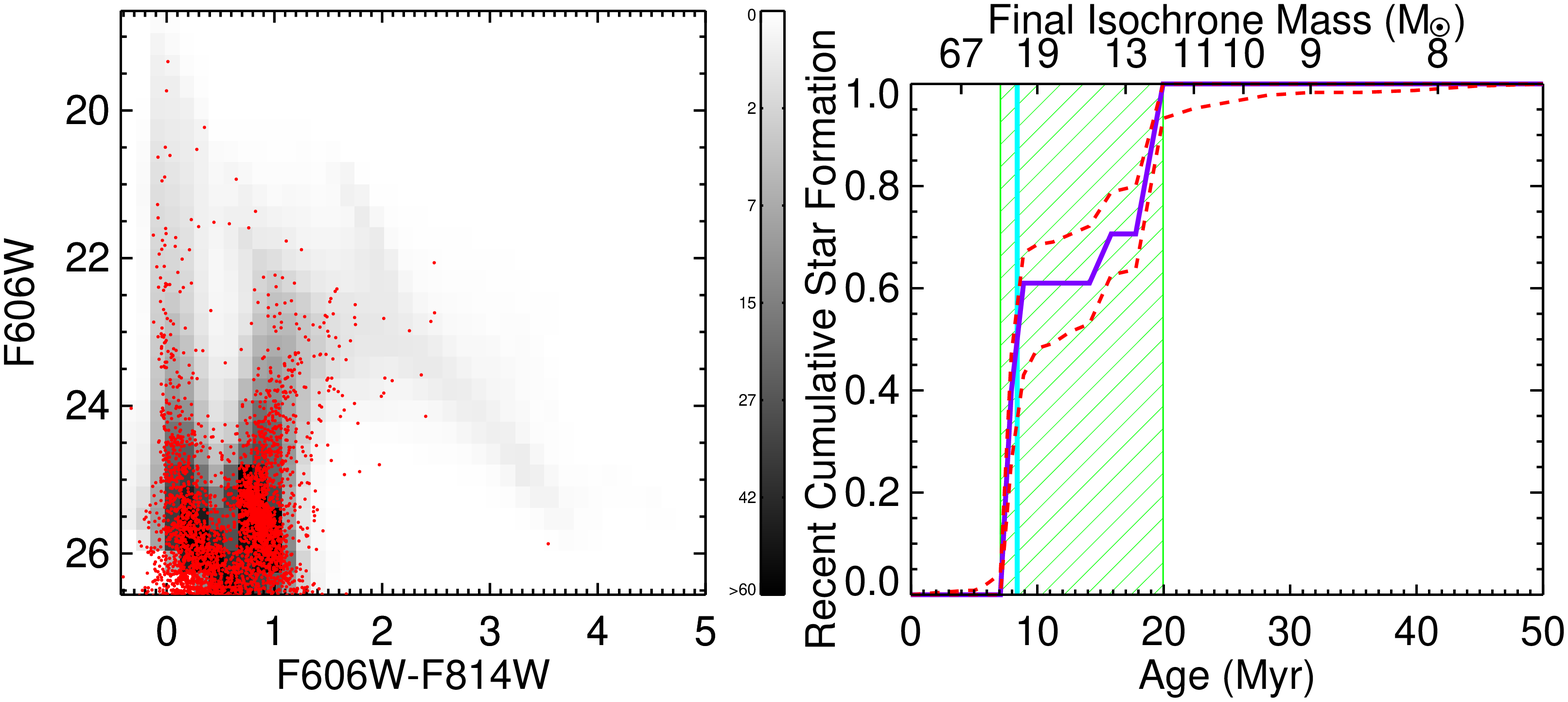,height=3.5in,width=7in}
\caption{Left Panel: Red points are observed F606W/F814W CMD for SNR
34 in M33. The background grayscale represents the best-fit model from
MATCH, with the stellar density scale on the right.  Right Panel:
Cumulative fractional SFH from the best-fit MATCH model over the most
recent 50 Myr.  The best fit SFH is the purple line. The orange
lines are 68\% uncertainties. The cross-hatched green region
represents the favored age and mass for this SNR progenitor, with
uncertainties calculated as described in \S2.4. The SFH is more spread
out than in the example in Figure~\ref{mass_dist1}, favoring a mass of
$24^{+4}_{-12}$. The blue line represents the median age.}
\label{mass_dist2}
\end{figure*}

\begin{table*}
\centering
\setlength{\extrarowheight}{3pt}
\begin{threeparttable}
\caption{Progenitor Mass Results for Newly Analyzed M31 SNR Sample}
\begin{tabular}{c c c c c c c}
\hline
SNR ID & \mzams & Progenitor Age & MS Stars & Total Stars & Stellar Mass Formed & Additional $dA_{\rm V}$ Applied \\
 & (\msun) & (Myr) & & & ($10^{2}$ \msun) & (Total $dA_{\rm V} - 0.5$) \\
\multicolumn{7}{c}{M31 SNR Progenitors} \\
\hline
2-038 & $8.4^{+0.2}_{-0.2}$ & $38^{+2}_{-2}$ & 191 & 4601 & 11 & 0.5\\
2-039 & $7.6^{+1.4}_{-0.3}$ & $46^{+4}_{-14}$ & 265 & 3940 & 25 & 0.5\\
2-040 & $7.5^{+0.2}_{-0.2}$ & $47^{+3}_{-3}$ & 272 & 5812 & 11 & 0.3\\
2-041 & $8.4^{+8.8}_{-0.2}$ & $37^{+2}_{-26}$ & 218 & 3666 & 31 & 0.7\\
2-042 & $17^{+1}_{-11}$ & $11^{+52}_{-1}$ & 187 & 4794 & 7 & 0.5\\
2-043 & $8.1^{+0.4}_{-0.8}$ & $40^{+10}_{-4}$ & 154 & 2177 & 12 & 0.3\\
2-052 & $8.8^{+0.2}_{-0.3}$ & $33^{+2}_{-2}$ & 272 & 4091 & 47 & 0.6\\
2-053 & $7.9^{+0.2}_{-0.2}$ & $42^{+2}_{-2}$ & 298 & 4126 & 22 & 0.8\\
2-054 & $118^{+INDEF}_{-110}$ & $2.2^{+37.6}_{-1.5}$ & 343 & 3950 & 1 & 0\\
BW45 & $8.5^{+17.5}_{-0.4}$ & $36^{+4}_{-29}$ & 265 & 5434 & 81 & 0.6\\
BW51 & $7.9^{+0.2}_{-0.2}$ & $42^{+2}_{-2}$ & 207 & 4206 & 11 & 0.5\\
BW58 & $8.6^{+8.6}_{-0.4}$ & $36^{+4}_{-24}$ & 264 & 4993 & 18 & 0.3\\
BW59 & $8.6^{+0.5}_{-0.9}$ & $35^{+9}_{-4}$ & 298 & 4624 & 56 & 0.7\\
BW63 & $15^{+28}_{-6}$ & $14^{+26}_{-9}$ & 335 & 5009 & 38 & 0.1\\
BW64 & $13^{+4}_{-2}$ & $16^{+6}_{-5}$ & 384 & 5089 & 46 & 0.1\\
BW79 & $7.6^{+4}_{-0.3}$ & $46^{+4}_{-26}$ & 332 & 4611 & 9 & 0.3\\
BW94 & $8.4^{+0.6}_{-0.3}$ & $37^{+3}_{-6}$ & 326 & 4343 & 53 & 0.7\\
BW95 & $8^{+9.1}_{-0.3}$ & $41^{+3}_{-30}$ & 390 & 4568 & 42 & 0.6\\
BW96 & $7.9^{+2.4}_{-0.5}$ & $43^{+7}_{-18}$ & 451 & 3709 & 47 & 0.1\\
BW98 & $14^{+3}_{-4}$ & $15^{+13}_{-4}$ & 204 & 4239 & 11 & 0.4\\
BW100 & $8.4^{+0.2}_{-0.2}$ & $38^{+2}_{-2}$ & 332 & 4940 & 8 & 0.3\\
BW101 & $8.2^{+0.4}_{-0.4}$ & $40^{+5}_{-4}$ & 271 & 4822 & 40 & 0.4\\
K293 & $8.8^{+INDEF}_{-0.7}$ & $34^{+6}_{-31}$ & 525 & 5682 & 47 & 0.4\\
K638 & $7.9^{+0.2}_{-0.2}$ & $42^{+2}_{-2}$ & 316 & 4749 & 7 & 0.3\\
K763 & $9.2^{+0.5}_{-1.8}$ & $31^{+19}_{-3}$ & 288 & 3650 & 16 & 0.2\\
K774 & $8.4^{+0.2}_{-0.2}$ & $38^{+2}_{-2}$ & 270 & 4694 & 8 & 0.1\\
K782 & $7.9^{+0.2}_{-0.2}$ & $42^{+2}_{-2}$ & 294 & 4616 & 17 & 0.3\\
K817 & $8^{+5.4}_{-0.3}$ & $41^{+3}_{-26}$ & 698 & 5068 & 38 & 0.3\\
K854A & $14^{+1}_{-6}$ & $16^{+34}_{-1}$ & 366 & 4194 & 18 & 0.6\\
K413 $^{\rm \bf A}$ & - & - & 338 & 5364 & 0 & 0.1\\
\hline
\end{tabular}
\begin{tablenotes}[para,flushleft]
\textbf{A:} K413 features no recent star formation. We interpret it as a likely Ia candidate and remove
it from our combined mass distributions in subsequent analysis. 
\end{tablenotes}
\end{threeparttable}
\end{table*}

\begin{table*}
\centering
\setlength{\extrarowheight}{3pt}
\begin{threeparttable}
\caption{Progenitor Mass Results for M33 SNR Sample}
\begin{tabular}{c c c c c c c}
\hline
SNR ID & \mzams & Progenitor Age & MS Stars & Total Stars & Stellar Mass Formed & Additional $dA_{\rm V}$ Applied \\
 & (\msun) & (Myr) & & & ($10^{2}$ \msun) & (MATCH -$dA_{\rm V}$ Flag) \\
\multicolumn{7}{c}{M33 SNR Progenitors} \\
\hline
20 & $8.6^{+0.3}_{-0.3}$ & $38^{+2}_{-2}$ & 1515 & 2816 & 44 & 0.2\\
30 & $8.2^{+14.4}_{-0.7}$ & $42^{+8}_{-33}$ & 1049 & 3014 & 91 & 0.6\\
34 & $24^{+4}_{-12}$ & $8.4^{+11.5}_{-1.3}$ & 1136 & 3024 & 150 & 0.8\\
37 & $10^{+1}_{-2}$ & $27^{+17}_{-2}$ & 368 & 1242 & 7 & 0\\
38 & $11^{+1}_{-4}$ & $25^{+32}_{-2}$ & 110 & 354 & 8 & 0\\
40 & $7.7^{+3.8}_{-0.3}$ & $47^{+4}_{-24}$ & 193 & 1008 & 51 & 0\\
41 & $9.5^{+18.8}_{-2.5}$ & $31^{+25}_{-24}$ & 1351 & 3317 & 120 & 0.6\\
43 & $16^{+3}_{-8}$ & $14^{+31}_{-3}$ & 188 & 1004 & 100 & 0\\
44 & $14^{+1}_{-7}$ & $16^{+29}_{-2}$ & 244 & 836 & 22 & 0\\
48 & $11^{+130}_{-1}$ & $26^{+2}_{-24}$ & 809 & 2805 & 72 & 0\\
49 & $9^{+0.5}_{-0.6}$ & $35^{+5}_{-3}$ & 1020 & 3081 & 78 & 0.3\\
50 & $19^{+1}_{-11}$ & $11^{+39}_{-1}$ & 125 & 628 & 74 & 0\\
57 & $8.7^{+35.8}_{-1.2}$ & $37^{+13}_{-32}$ & 824 & 4082 & 79 & 0.1\\
58 & $21^{+2}_{-12}$ & $9.8^{+25.7}_{-0.9}$ & 394 & 1170 & 160 & 0\\
59 & $12^{+1}_{-5}$ & $20^{+25}_{-2}$ & 905 & 3439 & 167 & 0.1\\
61 & $15^{+INDEF}_{-7}$ & $16^{+35}_{-14}$ & 255 & 927 & 164 & 0\\
62 & $19^{+2}_{-7}$ & $11^{+11}_{-1}$ & 428 & 1470 & 114 & 0\\
63 & $7.7^{+0.7}_{-0.2}$ & $47^{+3}_{-7}$ & 353 & 1313 & 35 & 0\\
64 & $7.7^{+0.2}_{-0.2}$ & $47^{+3}_{-3}$ & 575 & 3559 & 35 & 0.1\\
66 & $8.1^{+0.2}_{-0.2}$ & $42^{+2}_{-2}$ & 352 & 1511 & 63 & 0\\
67 & $7.8^{+1.1}_{-0.3}$ & $46^{+4}_{-11}$ & 454 & 1938 & 66 & 0\\
69 & $8.2^{+6.2}_{-0.3}$ & $41^{+3}_{-26}$ & 1589 & 2090 & 161 & 0.3\\
71 & $11^{+12}_{-1}$ & $26^{+2}_{-17}$ & 1506 & 2639 & 133 & 0.1\\
74 & $8.7^{+46.1}_{-0.3}$ & $37^{+3}_{-33}$ & 580 & 1924 & 58 & 0\\
75 & $9.9^{+7.2}_{-0.4}$ & $30^{+2}_{-17}$ & 134 & 1063 & 118 & 0\\
76 & $8.7^{+1.4}_{-0.4}$ & $37^{+3}_{-8}$ & 705 & 2359 & 71 & 0\\
77 & $8.7^{+INDEF}_{-0.3}$ & $37^{+3}_{-34}$ & 1145 & 2049 & 89 & 0.1\\
80 & $8.6^{+0.3}_{-0.3}$ & $38^{+2}_{-2}$ & 864 & 4120 & 84 & 0.2\\
81 & $14^{+6}_{-4}$ & $16^{+12}_{-6}$ & 1905 & 2372 & 243 & 0\\
83 & $16^{+101}_{-8}$ & $13^{+22}_{-11}$ & 1444 & 2221 & 391 & 0.2\\
84 & $113^{+INDEF}_{-103}$ & $2.7^{+25.5}_{-1.9}$ & 68 & 315 & 10 & 0\\
91 & $10^{+1}_{-1}$ & $28^{+3}_{-3}$ & 1217 & 1816 & 29 & 0\\
134 & $8.6^{+0.2}_{-0.3}$ & $37^{+2}_{-2}$ & 349 & 874 & 14 & 0\\
\hline
\end{tabular}
\begin{tablenotes}[para,flushleft]
\end{tablenotes}
\end{threeparttable}
\end{table*}

The uncertainties on our age and mass measurements are calculated as
described in \S2.4. Note that for progenitors with a very precise age
(e.g. 2-038), the quoted uncertainties are underestimated as our
age-to-mass conversion uncertainty likely dominates. However, as it is
not clear how to quantify this uncertainty, we do not include it in
our quoted range. Progenitors with upper errors of "INDEF" have upper
errors which exceed the mass range over which we can measure
($>52$~\msun~for M31 and $>55$~\msun~for M33). As a result, we may only apply lower limits for the
masses of these progenitors. Indeed, three progenitors have
\textit{median} masses above our upper limit. We include these values
for completeness in the table, but in truth the median mass is not
meaningful in this case, at least in terms of assigning a mass to the
progenitor star.

\begin{figure*}
\centering
\epsfig{file=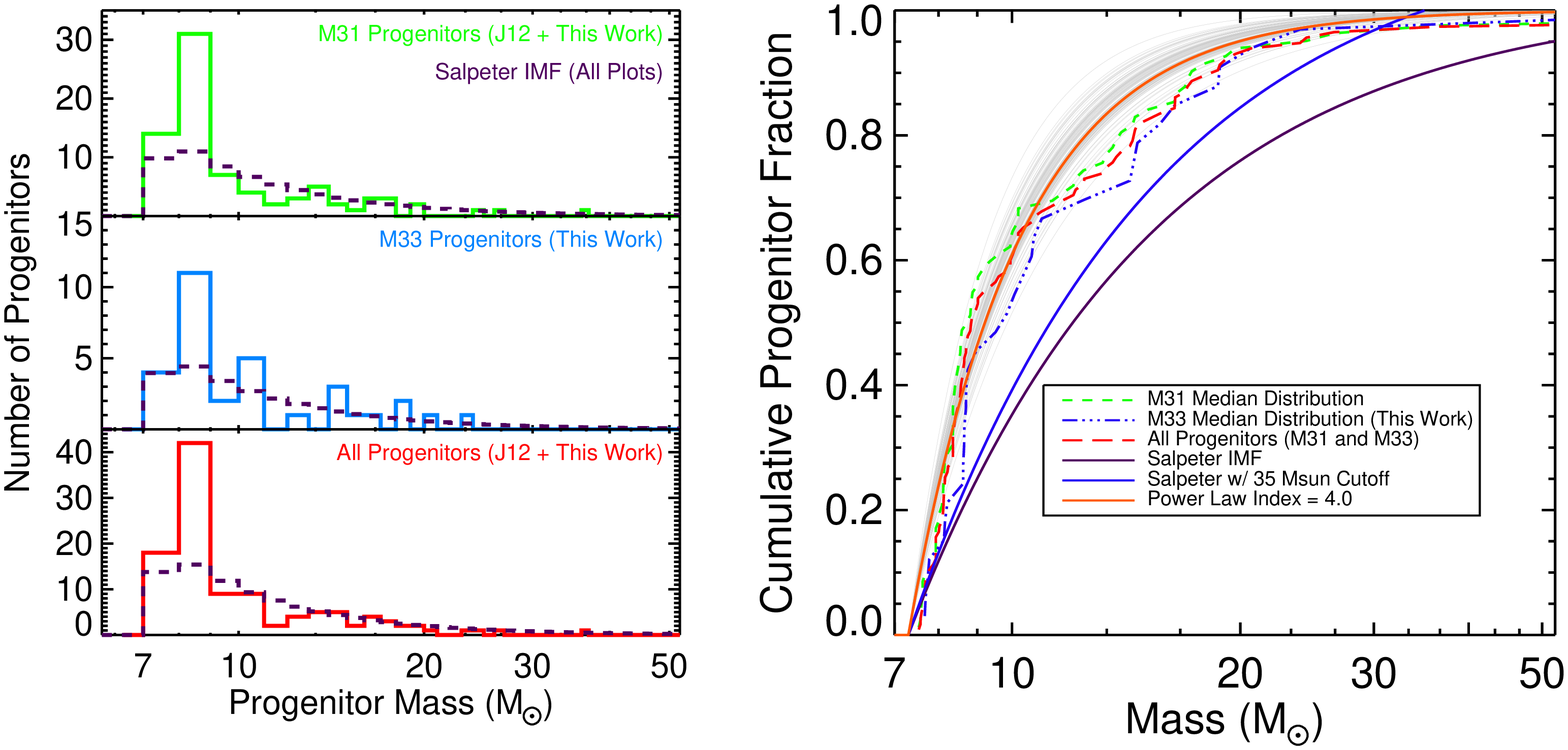,height=3.5in,width=7in}
\caption{Left Panels: Histogram of median progenitor
masses for all M31 progenitors (this work and J12), 
new M33 progenitors (this work), and all progenitors from
both galaxies. Histograms which would be expected from a Salpeter ($\alpha=2.35$)
IMF are plotted in dashed yellow lines. The Salpeter histograms are
normalized such that the total number of progenitors of the Salpeter distribution
is the same as that observed in each sample.
Right Panel: Cumulative distribution of
median progenitor masses for same distributions. We
also include a Salpeter IMF, a Salpeter IMF integrated only up to 35 \msun, and
a power-law IMF with slope $\alpha=4.0$. Note that all
median mass distributions are steeper than a Salpeter IMF.}
\label{mass_dist}
\end{figure*}

We present the collective progenitor mass distribution in two
ways. Fig.~\ref{mass_dist} shows a simple histogram of median
progenitor masses. We plot histograms for the separate M31 and M33
distributions, as well as the combined sample of all progenitors. We include the histograms
which would be expected to be observed if the SNe progenitor distribution followed
a Salpeter IMF, with the total number of stars normalized to be the same as that included
in the specified sample. 

The right panel of Fig.~\ref{mass_dist} displays the cumulative median mass
distributions for the above samples. In addition, we also plot 100 randomly selected
slopes from the MCMC fitting procedure (see \S4.1), showing the approximate
extent of the probability distribution for the mass function exponent. The highlighted orange line
is for $\alpha=4$. Note the apparent offset between the MCMC realizations of the function and the
median progenitor distributions. This is due to the fact that many of the more massive progenitors
have uncertainties extending back to significantly older ages, emphasizing the importance of properly
considering the uncertainties in the mass estimates.

We also include, for reference, two
Salpeter IMFs ($dN/dM \propto M^{-2.35}$), one integrated to 120 \msun~and one integrated to 35 \msun. 
Note that the measured distributions are all
visibly steeper (featuring fewer high mass progenitors) than would be
expected from a Salpeter IMF.

\section{QUANTIFYING THE PROGENITOR MASS DISTRIBUTIONS}
In this section, we present our analysis of the recovered progenitor
mass distributions.  We evaluate these distributions in two different
ways. First for comparison with J12, we apply the Kolmogorov-Smirnov
(K-S) test between our distributions and those of several power law
functions. Note that our use of the K-S test neglects uncertainties on our individual measurements,
so we include it purely to verify the consistency of our current results with our previous analysis.
Next, for reliable comparisons to models which properly
include our uncertainties, we use a more sophisticated maximum-likelihood technique
(adapted from \citealt{weisz2013}) to fit power laws to our progenitor
mass distributions. We use these maximum-likelihood inferred results for all scientific
interpretations.

We first use a K-S test to verify consistency of our results with J12.
Our procedure is to take the observed distribution
of median masses for the listed segment of our sample and perform a
1-sided K-S comparison to a sample IMF, defined as $dN/dM \propto
M^{-\alpha}$.  Values of $\alpha$ for which the K-S test returns a
$<5\%$ chance ($P<0.05$) of being from the same distribution are
defined as being inconsistent with the given mass distribution at 95\%
confidence. For all IMF comparisons, we integrate the model IMF from
7.3~\msun, the minimum mass value we found in J12, to 120~\msun.  The
exact selection of maximum mass value does not make a significant
difference because very massive stars contributed little to the
expected numbers.

The primary result from this analysis is that all of our sample
selections (only M31, only M33, and combined) are inconsistent with a
Salpeter IMF. The range of power-law indexes that are consistent with
our simple median mass distributions according to the K-S test
($P>0.05$) are 2.4--4.6 for M33, 3.8--4.8 for M31, and 3.7--4.3 for
the combined sample.  In other words, we have a paucity of massive
stars in our progenitor distributions.  We would expect to recover
more massive stars if they have been formed following a Salpeter IMF
and all explode as SNe. This finding is consistent with the main
result of J12 by expansion to a larger sample, and supports a growing
number of other lines of evidence suggesting that not all massive
stars explode as SNe.  However, because these comparisons do not take
into account the uncertainties of our mass estimates, they do not
represent a reliable characterization of the range of acceptable
power-law indexes.  Therefore, while it is encouraging to find
consistency between J12 and our new larger samples, we now turn to
more sophisticated fitting techniques to determine how well we can
characterize the distributions with our current measurements.

\subsection{MCMC Mass Distribution Fitting}

To properly include uncertainties in our mass distribution fitting, we
adopted a probabilistic framework developed by \citet{weisz2013} to
fit a power-law function to our mass distributions. A full probability distribution
for the entire sample is created from the individual mass measurements and uncertainties.
We then fit a power-law distribution using
the Markov Chain Monte Carlo (MCMC) sampler {\it emcee}
\citep{foreman-mackey2013} to sample the posterior distributions. We then extracted
the functional parameters and robust uncertainties from these posterior distributions.
Quoted uncertainties correspond
to the 16th and 84th percentiles of the probability distributions.
When performing our fitting, we set a prior that M$_{\rm max}{>}$20
M$_{\odot}$, and fixed the minimum mass to 7.3~\msun. As an illustration of the distribution
of power-law indices, we plot 100 randomly selected MCMC steps in Fig.~4.

We found that we were unable to constrain the maximum mass beyond our
prior if both the mass function slope and the maximum mass were left
as free parameters (all values $>$20~\msun were found to be consistent
with the data if ${\alpha}{\sim}4$).  Therefore, we used two different
approaches to fit our distributions. For one, we assumed the
distribution followed a power-law function of the form $dN/dM\propto
M^{-\alpha}$, with $\alpha$ left as a free parameter and with $M_{\rm
max}$ fixed to 90 \msun. We also performed an MCMC run in which we
fixed the functional form of the mass function to be a Salpeter IMF,
$dN/dM\propto M^{-2.35}$, as one may expect to recover if all massive
stars produce SNe. In these fits, $M_{\rm max}$ was constrained by the
data. \citet{weisz2013} explicitly notes that a meaningful limit on the
maximum mass cannot be
constrained by the data beyond simply returning the most massive star observed,
so we emphasize that fit values for $M_{\rm max}$ in particular should be considered
illustrative. The precise values are not necessarily meaningful, whereas
the fits for a free slope are more reliable. 

The results from these fits are listed in Table 5. We do not
attempt to evaluate which model description is a better fit to the
data. There is evidence that the regions of progenitor masses over
which SNe may explode may actually be quite complex
(e.g. \citealt{sukhbold2014}), making either parametrization of the
mass distribution a likely oversimplification. However, the result of
a paucity of massive progenitors is robust regardless of model
selection. The use of the MCMC technique properly incorporates
our measured uncertainties and rules out a Salpeter mass function
for our progenitor sample.

\setlength{\extrarowheight}{3pt}
\begin{table}
\begin{threeparttable}
\caption{MCMC Inferred Mass Distribution Parameters}
\begin{tabular}{c c c}
\hline
Sample & Best Fit $\alpha$ $^{\rm \bf A}$ & Maximum Mass (\msun)$^{\rm \bf A}$\\
\hline \\
\multicolumn{3}{c}{Index $\alpha$ as Free Parameter, $M_{\rm max}=90$ \msun}\\
\hline 
M31 Sample (J12 + This Work) & $\alpha=4.4^{+0.4}_{-0.4}$ & $M_{\rm max}=90$ \\
M33 Sample (This Work) & $\alpha=3.8^{+0.5}_{-0.4}$ & $M_{\rm max}=90$ \\
Full Sample (J12 + This Work) & $\alpha=4.2^{+0.3}_{-0.3}$ & $M_{\rm max}=90$ \\
\hline \\
\multicolumn{3}{c} {Index Fixed to Salpeter ($\alpha=2.35$), $M_{\rm max}$ as Free Parameter}\\
\hline 
Full M31 Sample (J12 + This Work) & 2.35 & $M_{\rm max} = 38^{+15}_{-6}$ \\
Full M33 Sample (This Work) & 2.35 &  $M_{\rm max} = 41^{+45}_{-12}$ \\
Full Distribution (J12 + This Work) & 2.35 &  $M_{\rm max} = 35^{+5}_{-4}$   \\
\hline
\end{tabular}
\begin{tablenotes}[para,flushleft]
\textbf{A:} Quoted uncertainties come from 16th and 84th percentiles of the probability distributions.
\end{tablenotes}
\end{threeparttable}
\end{table}

We find that, independent of the sample that we fit, all distributions
analyzed favor a model in which some fraction of the massive star
population do not explode as SNe, whether it is parametrized as a
very steep mass function or a Salpeter mass function with a cutoff.
Furthermore, the M31 and M33 fit values indicate that both the slope
and the maximum mass cutoff are consistent to within the
uncertainties, suggesting that any metallicity or star formation
intensity effects do not change the index by more than $\pm{\sim}$1.
While this range in power-law indices is still large, the distributions
are all still steeper than Salpeter regardless. A more precise comparison of the M31 and M33
distributions would naturally be interesting, but we are unable to make a more constraining
statement than this given the current size of the dataset and the uncertainties involved.

\subsection{A Possible High Mass Cutoff for Producing SN}

Both the individual M31 and M33 progenitor mass distributions, as well
as the full progenitor mass distribution, are steeper than a Salpeter
IMF with at least 95\% confidence, regardless of the method used to
fit the distribution. Thus, both the full sample and the individual
M31 and M33 distributions display a paucity of massive stars compared
to a simple model IMF.  If we assume that a Salpeter distribution is a
reasonable description of the massive star population, then this
paucity suggests that some fraction of massive stars are not in our
SNR progenitor sample. Recent theoretical and observational work has
suggested that in certain mass ranges, massive stars may not undergo
SN explosions (e.g. \citealt{smartt2009,horiuchi2011,dessart2011,sukhbold2014}).
Our observation is consistent with the interpretation that there are
certain mass ranges in the massive star population in which stars fail
to end their lives as CCSN.

\subsection{Potential Effects From Biases in the SNR Sample}
One important caveat is that the interpretation of an observed upper-mass
CCSNe cutoff relies on sampling an unbiased population of massive stars,
\textit{i.e.}~that our SNR catalogs are unbiased to any one type of progenitor star
environment.  Since we would expect higher-mass progenitors to be
preferentially found in more-extincted regions, this may cause SNR
from higher-mass progenitors to be systematically under-sampled in SNR
catalogs. We have now expanded the sample to include SNR from M33,
which is less inclined to our line of sight and lower metallicity
\citep{bresolin2011,zurita2012}.  Both of these qualities should
reduce the amount of extinction, somewhat reducing these biases.
However, we also note that a commonly-used method for identifying SNRs
is by their high [S~II]/H$\alpha$ flux ratios
\citep[e.g.][]{gordon1998}. The youngest regions will have more
photoionized HII, increasing the H$\alpha$ flux and making SNR
identification more difficult.  We made some attempt to reduce this
bias by incorporating SNR catalogs that use different methodology.  In
particular, the \citet{long2010} M33 SNR catalog incorporates
observations from radio to X-ray to catalog SNRs.
In any case, given
that SNRs do not stand out if they are located within a photo-ionized
ISM, we are currently unable to definitively distinguish between the
possibility of SNRs from more-massive progenitors being more difficult
to identify, and very high-mass stars not producing SNe.

In addition to difficulties in detection due to obscuration, any
inherent differences in luminosity or lifetime of SNRs as a function of
progenitor mass could also introduce some bias. These properties
are dependent upon both the energy of the SN and the density of the medium
into which the SNR expands.
If one would appeal to this as an explanation
for the lack of high mass progenitors, this would imply that SNR from higher mass
stars have inherently lower luminosities and/or inherently shorter lifetimes. 
At present, we have no
real ability to test either of these cases with our methodology. If there are significant
differences in SNR frequencies or properties as a function of progenitor mass, then
naturally we are no longer dealing with an unbiased sample of all exploded
massive stars, and our conclusions about the SNe progenitor mass distribution
are weakened.

\section{SUMMARY}
In this work, we estimated progenitor masses of 33 SNR in M33 and 29
SNR in M31 using identical methodology to \citet{jennings2012}. After
combining these with the 53 CCSNe progenitor mass estimates from J12,
we constrain the progenitor distributions with a probabilistic
technique which incorporates the uncertainties on the individual
progenitor mass estimates. We find that the progenitor mass
distributions of M31 and M33 both display a paucity of massive stars
when compared to a Salpeter IMF. This suggests that some fraction of
massive stars are not exploding as SNe, a similar finding to that seen
in other theoretical and observational work
\citep{smartt2009,horiuchi2011,dessart2011,kochanek2014}.  Our work
represents an independent and complementary technique to these other
methods. Now that the result has been expanded to M33, we have
verified that this result holds across a range of galaxy inclination,
metallicity, and morphology.  However, we note that due to potential
biases in the SNR catalogs, we cannot rule out the possibility that
the SNR surveys have preferentially missed the SNRs of more massive
SN progenitors.

In addition to this main result, we have also investigated other
interesting properties of the progenitor mass distributions. Assuming
a Salpeter IMF is the expected distribution if all massive stars
produce SNe, we estimated the upper-mass cutoff necessary to be
consistent with our data. The data from both galaxies are consistent
with a maximum mass for core-collapse SNe of ${\sim}35-45$ \msun.  We
also compare the progenitor mass distributions of M33 and M31, finding
them to be consistent with one another, albeit to within large uncertainties.
At this stage, we believe we have examined all available archival
\textit{HST} data of sufficient depth to apply our stellar population
analysis techniques in M31 and M33. We will be unable to offer a more precise
comparison of the M31 and M33 SNR populations without additional observations
of SNR in either galaxy.

\acknowledgements
Z.G.J. and B.F.W. are supported in part by AR-12834. B.F.W. is also supported
in part by GO-12055. Z.G.J. is supported in part by a National Science Foundation Graduate
Research Fellowship.
This work is based on observations made with the NASA/ESA Hubble Space Telescope,
obtained from the data archive at the Space Telescope Science Institute.
Support for this work was provided by NASA through Hubble Fellowship grant 51273.01 awarded
to K.M.G. by the Space Telescope Science Institute.
STScI is operated by the Association of Universities for Research in
Astronomy, Inc. under NASA contract NAS 5-26555.
Support for DRW is provided by NASA through Hubble Fellowship grant HST-HF-51331.01 awarded by the Space Telescope Science Institute.

%\bibliographystyle{apj}
%\bibliography{bibtex.bib}

\end{document}